\documentclass{aastex}
\usepackage{emulateapj5,psfig}

\newcommand{\etal}{{\it\ et\,al.\ }}
\newcommand{\beq}{\begin{equation}}
\newcommand{\eeq}{\end{equation}}

\newcommand{\microk}{$\mu{\mbox{K}}$}

\begin{document} 
\title{A High Spatial Resolution Analysis of the MAXIMA-1 Cosmic
Microwave Background Anisotropy Data}

\author{A.~T.~Lee\altaffilmark{1,2,3}, P.~Ade\altaffilmark{4},
A.~Balbi\altaffilmark{2,3,5}, J.~Bock\altaffilmark{6,7},
J.~Borrill\altaffilmark{3,8}, A.~Boscaleri\altaffilmark{9}, \\P.~de
Bernardis\altaffilmark{10}, P.~G.~Ferreira\altaffilmark{11,12},
S.~Hanany\altaffilmark{3,13}, V.~V.~Hristov\altaffilmark{7},
A.~H.~Jaffe\altaffilmark{3}, P.~D.~Mauskopf\altaffilmark{14},
C.~B.~Netterfield\altaffilmark{15}, E.~Pascale\altaffilmark{9},
B.~Rabii\altaffilmark{1,2,3,16}, P.~L.~Richards\altaffilmark{1,3},
G.~F.~Smoot\altaffilmark{1,2,3,16},
R.~Stompor\altaffilmark{3,16,17}, C.~D.~Winant\altaffilmark{1,3,16},
J.~H.~P. ~Wu\altaffilmark{18} }

\altaffiltext{1}{Dept. of Physics, University of California, Berkeley
CA, USA}
\altaffiltext{2}{Division of Physics,Lawrence Berkeley National Laboratory,
Berkeley, CA, USA}
\altaffiltext{3}{Center for Particle Astrophysics, University of
California, Berkeley, CA, USA}
\altaffiltext{4}{Queen Mary and Westfield College, London, UK}
\altaffiltext{5}{Dipartimento di Fisica, Universit\`a Tor Vergata,
Roma, Italy}
\altaffiltext{6}{Jet Propulsion Laboratory, Pasadena, CA, USA}
\altaffiltext{7}{California Institute of Technology, Pasadena, CA,
USA}

\altaffiltext{8}{National Energy Research
Scientific Computing Center, Lawrence Berkeley National Laboratory,
Berkeley, CA, USA}
\altaffiltext{9}{IROE-CNR, Firenze, Italy}
\altaffiltext{10}{Dipartimento di Fisica, Universit\`a La Sapienza,
Roma, Italy}
\altaffiltext{11}{Astrophysics, University of Oxford, Oxford, UK}
\altaffiltext{12}{CENTRA, Instituto Superior Tecnico, Lisboa, Portugal}
\altaffiltext{13}{School of Physics and Astronomy,
University of Minnesota/Twin Cities, Minneapolis, MN, USA}

\altaffiltext{14}{Department of Physics and Astronomy, University of
Wales, Cardiff, UK}
\altaffiltext{15}{Dept. of Physics and Astronomy, University of
Toronto, Canada}
\altaffiltext{16}{Space Sciences Laboratory, University of Californin,
Berkeley, CA, USA}
\altaffiltext{17}{Copernicus Astronomical Center, Warszawa, Poland}
\altaffiltext{18}{Department of Astronomy, University of California,
Berkeley, CA, USA}

\begin{abstract}

We extend the analysis of the MAXIMA-1 cosmic microwave background
(CMB) data to smaller angular scales.  MAXIMA, a bolometric balloon
experiment, mapped a 124 deg$^2$ region of the sky with 10\arcmin\
resolution at frequencies of 150, 240 and 410 GHz during its first
flight.  The original analysis, which covered the multipole range $36
\leq \ell \leq 785$, is extended to $\ell = 1235$ using data from
three 150 GHz photometers in the fully cross-linked central 60 deg$^2$
of the map.  The main improvement over the original analysis is the
use of 3\arcmin\ square pixels in the calculation of the map.  The new
analysis is consistent with the original for $\ell < 785$.  For $\ell
> 785$, where inflationary models predict a third acoustic peak, the
new analysis shows power with an amplitude of $56 \pm 7$ \microk\ at
$\ell \simeq 850$ in excess to the average power of $42 \pm 3$
\microk\ in the range $441 < \ell < 785$.

\end{abstract}


\section{Introduction} 

The power spectrum of temperature fluctuations in the cosmic microwave
background (CMB) has recently been measured over a broad range of
angular scales \citep{debernardis00, hanany_00}. The results are being
used to discriminate between cosmological models and to determine
cosmological parameters \citep{lange00, balbi_00, dodelson_knox00,
tegmark_zald2000, pierpaoli_2000,jaffe_maxiboom}. There is strong
evidence for a first peak in the power spectrum at $\ell \sim 200$
which is consistent with a flat universe in adiabatic inflationary
models with cold dark matter.  Fluctuations were also detected in the
multipole range $400 \le \ell \le 785$ at smaller scales than those of
the first peak.

MAXIMA is a balloon-borne experiment optimized to map the CMB
anisotropy over a broad range of angular scales from 10\arcmin\ to
5\degr.  The first flight, MAXIMA-1, observed 124 square degrees of
the sky.  We have published an analysis of this data over the
spherical harmonic multipole range $36 \leq \ell \leq 785$
\citep{hanany_00}. In this paper, we extend the original analysis to
$\ell = 1235$ using the data from three 150 GHz photometers.  These
measurements are a powerful discriminant between cosmological models
and allow parameters to be determined with better accuracy.  A
companion paper \citet{aardvark_00} discusses the cosmological
significance of these results.

The MAXIMA instrument is described in detail by \citet{lee_instrument}
and \citet{hanany_00}.  In sections 2-4, we describe the analysis of
the telescope beams, the pointing reconstruction, and the estimation
of instrument noise, which are especially important for the
high-$\ell$ region of the power spectrum.  In section 5, we present
the map and the angular power spectrum.  In sections 6 and 7, we show
that foreground contributions to the data are negligible and describe
tests for systematic errors.
 
\section{Telescope Beams}

For a total-power radiometer such as MAXIMA, a finite telescope beam
size introduces an exponential reduction in sensitivity at
high-$\ell$.  Extending the MAXIMA power spectrum to higher $\ell$
values thus depends on an accurate knowledge of the beam patterns.
Each of the 16 elements of the MAXIMA bolometer array were designed to
have a FWHM Gaussian beam size of 10\arcmin, which reduces power by a
factor of five at $\ell \sim 1000$.  The beams were measured in flight
using observations of Jupiter, which has an angular diameter of
0.5\arcmin.  Observations were carried out at fixed elevation with the
primary mirror scanning $\sim 4$\degr\ peak-peak in azimuth at 0.45 Hz
and the gondola azimuth fixed.  Sky rotation gave a raster map of
Jupiter with more than 200 mirror scans across each array element.
The reconstructed beam profiles have a peak signal-to-noise ratio of
more than 1000.  The beams are close to circular near the peak, but
have a moderate asymmetry in the wings (see Fig. 9 in
\citet{wu_etal00}).

\citet{wu_etal00} have developed a formalism for calculating the
effects of asymmetric beams.  We find that data measured with slightly
asymmetric beams can be analyzed by using an effective symmetric beam.
This formal treatment has been tested using simulated observations of
a CMB sky with the observation strategy and measured beams for
MAXIMA-1.  Because of the high degree of symmetry near the beam
centers, the estimated error in $C_\ell$ introduced by this technique is
less than 1\% at $\ell = 1000$.

The uncertainty in the measured beam profiles is a source of error at
high $\ell$.  There is a 5\% uncertainty in the measurement of the
long and short axis FWHM, which is largely due to uncertainty in the
baseline subtraction for each of the one-dimensional Jupiter scans.
There are minor contributions to this error from uncertainties in
pointing reconstruction and the measured bolometer time constant.
This error gives an error in $C_\ell$ that increases with $\ell$
\citep{wu_etal00}.  The error is 6\% at $\ell = 500$ and 17\% at $\ell
= 1000$.  The errors in the $\ell$ bins are correlated, so the power
spectrum will be biased either up or down with an amplitude that
increases with $\ell$.  

\section{Pointing Reconstruction}

	The MAXIMA experiment observes each sky pixel repeatedly to
achieve an adequate signal-to-noise ratio.  Errors in telescope
pointing reconstruction smear the signal between pixels.  

There are four temporal modulations of the beam in the MAXIMA
experiment.  The primary mirror scans in azimuth by $\sim 4$\degr\
peak-peak at $\sim 0.45$ Hz, the gondola scans in azimuth by $\sim
8$\degr\ peak-peak at $\sim 20$ mHz, sky rotation is used to scan in
right ascension, and we map the same sky region twice during the
flight at different elevations to achieve cross linking of pixels.
Information from several sensors is combined to give the telescope
pointing as a function of time, and each sensor contributes to the
total pointing error.

The dominant pointing error comes from the primary mirror motion, as
the mirror is rotated left or right.  The elevation angle of the beams
increases as the square of the azimuth scan angle. This elevation
change can approach several arc minutes.  We measure the centering of
the primary mirror motion during flight with a $\sim 1\degr$
uncertainty, which gives a 0.8\arcmin\ elevation uncertainty.

A CCD camera mounted to the inner frame of the gondola measures the
attitude of the telescope by monitoring stars.  A dedicated processor
calculates the position of the two brightest stars in the field of the
camera.  By comparing the reconstructed and known separations of the
stars, we estimate a 0.45\arcmin\ rms error in gondola orientation.

The total pointing reconstruction error is 0.95\arcmin\ rms, including
the described contributions from the primary mirror position and the gondola
attitude errors and several additional minor contributions.  Although some of
the pointing errors are systematic rather than random, the multiple
modulations in MAXIMA's observation strategy tend to randomize the
errors when projected on the sky.  Therefore, we model the pointing
error as a simple Gaussian distributed error.

We simulate an observation with such a Gaussian pointing
reconstruction error to assess the effect on the measured power
spectrum.  As expected, the main effect is a systematic underestimate
of power at high $\ell$.  In principle, this bias is predictable and
can be accounted for accurately.  However, since the pointing error is
not perfectly Gaussian, we conservatively assign a symmetric random
error equal to the size of the bias which is $\delta C_\ell/C_\ell$
$\sim 10\%$ at $\ell = 1000$.  The $C_\ell$ error for each $\ell$ bin
due to pointing reconstruction error is given in
Table~\ref{tab:cl_estimates}.  As with the beam shape uncertainty,
this fully correlated error has the effect of increasing or decreasing
power monotonically with increasing $\ell$.

\section{Noise Estimation}

The data contain both the signal from the sky and noise from the detector
system.  The procedures for map making and power spectrum estimation
require an accurate estimate of the detector system noise to correctly
estimate the signal.  The accuracy of the noise estimation is most
critical in the high-$\ell$ portion of the power spectrum, where the
data become increasingly noise dominated.  This trend occurs because
$C_\ell$ falls roughly as $\ell^2$ and because the telescope beam
reduces the signal at high $\ell$.

As described in \citep{hanany_00}, we observe a spurious
primary-mirror synchronous signal which complicates the noise
estimation because it is larger than the rms amplitude of the noise.
We use two methods to estimate the detector noise.  In the first
method, we assume that noise dominates the signal at all frequencies
other than the primary mirror modulation frequency and its harmonics.
The noise at those frequencies is calculated by interpolating data
from the power spectrum at nearby frequencies.  In the second method, we use
the iterative noise estimation procedure of
\citet{ferreira_jaffe_2000}.  Here, the noise and signal are both
estimated iteratively in the map-making stage.  We iterate the noise
using a map with 8\arcmin\ pixels to limit computation time.  This
resolution should be sufficient, since the signal power decreases at
smaller angular scales.

We make 3\arcmin-resolution maps using the two noise estimates.  We
calculate the power spectrum for $\sim 23,000$ pixels in the
``central'' region of the maps, and we find that the difference
between the two spectra is much less than the statistical error for
all $\ell$ bins.  This central region covers roughly half the area of
the map (60 deg$^2$) where the observations are fully cross linked,
the sampling is the most uniform, and the signal-to-noise ratio is
highest.  In this way, we reduce both the time required to search for
systematic errors and the computation time.  This procedure increases
statistical error bars by $< 10\%$ for $\ell > 500$ and by $< 3\%$ at
$\ell = 1000$.

The power spectrum of the difference between maps from two photometers
in the array is a stringent test of the noise estimation process.  A
mis-estimate of the white noise level of the detector system gives a
power spectrum that diverges from zero as $\ell^2$.  Correlation in
the noise between detectors can give the same behavior.  In a separate
analysis we do not find significant correlation between the time
streams in the 0.1 - 30 Hz frequency range used in the analysis aside
from that caused by the spurious chopper-synchronous signal.  This
signal is removed during the map making stage resulting in maps
without correlations.  We also find that histograms of the
temperatures in the 8\arcmin\ difference maps are consistent with the
distributions expected for no noise correlations at a
Kolmogorov-Smirnov significance level larger than 10\%.  As discussed
in section 7, the power spectra obtained using difference maps are
consistent with zero, with no apparent systematic divergence at high
$\ell$.

\section{Map and Angular Power Spectrum}
\label{sec:map_and_power_spectra}

The map-making and power spectrum analysis procedures are similar to
those described by \citet{hanany_00}.  In that analysis we used
5\arcmin\ $\times$ 5\arcmin\ pixels to limit computation time, and
we limited the analysis to an $\ell$ range where such pixelization does
not bias the results.  In principle, the smearing effect of larger
pixels can be accounted for by deconvolving a pixel window function.
In practice, however, this procedure may introduce a systematic bias
in the power spectrum due to the uneven spatial distribution of
samples in each pixel.  We choose a 3\arcmin\ square pixel for the
present analysis, since this pixel size does not compromise the
resolution of the map \citep{wu_etal00}.

In this paper, we report on the analysis of data from three 150 GHz
photometers using a 3\arcmin\ pixelized map. \citet{hanany_00}
analyzed the data from three 150 GHz photometers and a 240 GHz
photometer using 5\arcmin\ pixelization.  The data from all four
photometers passed all tests for consistency up to $\ell = 785$.
In this work with 3\arcmin\ pixels, we exclude the 240 GHz data
because it does not pass consistency tests above $\ell = 785$.  The
power spectrum of difference maps between the 240 GHz data and any 150
GHz data are inconsistent with zero in this high $\ell$ region.  One
possible reason for this discrepancy is that the spurious
primary-mirror synchronous signal is larger by a factor $> 2$ in the
240 GHz data compared to the 150 GHz data.

The raw data preparation is described in \citet{hanany_00}.  We remove
data during calibration events, and we also remove short sections of
data which deviate by $> 4~\sigma$ such as cosmic ray events and
telemetry drop-outs.  We deconvolve the transfer functions of the
bolometers and readout electronics and estimate the noise power
spectrum from sections of the time stream that had no long gaps
(Stompor et al., in preparation).  We marginalize over frequencies
lower than 0.1 Hz and higher than 30 Hz, where we do not expect
appreciable optical signals.

The data are calibrated using observations of the CMB dipole, and
periodic optical stimulator events are used to account for a small
drift in calibration during the flight \citep{hanany_00}.  The
calibrated time stream data, the pointing solution, and the estimate
of the noise power spectrum are combined to produce a maximum
likelihood pixelized map of temperature anisotropy and a pixel-pixel
noise correlation matrix for each photometer
\citep{wright_96,tegmark_97,bond_etal99} using the MADCAP software
package \citep{borrill_1999}.

A combined maximum-likelihood temperature anisotropy map is produced
by adding individual maps with a weight inversely proportional to
their noise correlation matrices. This map, shown in
Figure~\ref{fig:maxima1_map}, contains $\sim 40,000$ 3\arcmin\ square pixels.
We assign a calibration uncertainty of 4\% to the magnitude of
temperature fluctuations in the combined map \citep{hanany_00}.

We calculate the angular power spectrum $C_{\ell}$ of the central
region of the combined map using the MADCAP \citep{borrill_1999}
implementation of the maximization of the likelihood function
following \citet{bjk98}.  The MADCAP implementation assumes that the
beam shape has axial symmetry.  We produce an effective beam for the
analysis of the combined map by noise-weighted averaging the
individual beams. As discussed in section 2, we follow the procedure
of \citet{wu_etal00} to find a symmetric approximation for the
effective beam.  We apply no corrections for the negligible ($< 1\%$ at
$\ell = 1000$) smoothing of the 3\arcmin\ pixels.

We calculate the power spectrum of the temperature fluctuations using
15 bins in $\ell$ over the range $ 3 \la \ell \la 1800 $ assuming a
constant $\ell(\ell +1)C_{\ell}$ band power in each bin, and
marginalizing over the CMB monopole and dipole.

The benefits of only using the central region of the map for power
spectrum analysis are discussed in section 4.  The penalty is an
increased error that falls with increasing $\ell$ due to the rise in
sample variance.  To minimize this effect, we create a composite power
spectrum which uses the $\ell < 335$ points from the original
\citet{hanany_00} analysis.  We choose this transition point to limit
the maximum error bar increase to $\sim 20\%$.  For bins with $\ell <
335$, we marginalize over all the higher $\ell$ bins from
\citet{hanany_00}.  For the bins with $\ell > 335$ obtained using the
central region, we marginalize over the bins at $\ell < 335$ and $\ell
> 1235$. For both $\ell$ regions, we diagonalize the $\ell$-bin
correlation matrix using a variant of techniques discussed in
\citet{bjk98}.  The correlations between the dominant bin and adjacent
bins are typically less than 10\%.

Table \ref{tab:cl_estimates} lists the dominant bins, the $C_{\ell}$
estimates, and the $\Delta T = \sqrt{\ell(\ell +1)C_{\ell}/2\pi}$
estimates for the corresponding uncorrelated linear combinations of
bins.  We quote $1 \sigma$ errors on the $C_{\ell}$ estimates assuming
68\% confidence intervals using the offset log-normal distribution
model of \citet{bjk00}.  These statistical errors do not include three
additional independent sources of systematic uncertainty, which are
fully correlated between $\ell$ bins.  The $1\sigma$ calibration error
is a constant 8\% of $\ell(\ell+1)C_{\ell}/2\pi$.  The
$\ell$-dependent errors due to the beam-diameter uncertainty and
pointing reconstruction uncertainty are given in
Table~\ref{tab:cl_estimates}. Information on the shape of the
bin-power likelihood functions and window functions will be made
available on the MAXIMA web site (http://cfpa.berkeley.edu/maxima).
\begin{table*}
\begin{center}
\begin{tabular}{cclccl} \hline \hline
$\ell_{eff}$ & $[\ell_{min},\ell_{max}]$ & $\ell(\ell+1)C_{\ell} /2\pi$ & Beam Error & Pointing Error &$\Delta T$ \\
   &         & ($\mu K^{2}$) & (\%) & (\%) & (\microk) \\ \hline 

77 & [ 36, 110]& $1999_{-506}^{+675}$ & $\pm 0$ & $\pm 0$  &          $45_{-6}^{+7}$ \\
147 & [ 111, 185]& $2960_{-554}^{+682}$ &  $\pm 0.6$ &  $\pm 0.2$ &        $54_{-5}^{+6}$ \\
222 & [ 186, 260]& $6092_{-901}^{+1052}$ &  $\pm 1.5$ &  $\pm 0.4$ &       $78_{-6}^{+6}$ \\
294 & [ 261, 335]& $3830_{-577}^{+670}$ &  $\pm 2.5$ &  $\pm 0.8$ &        $62_{-5}^{+5}$ \\
381 & [ 336, 410]& $2270_{-471}^{+569}$ & $\pm 3.5$ &  $\pm 1.2$ &        $48_{-5}^{+6}$ \\
449 & [ 411, 485]& $1468_{-325}^{+387}$ & $^{+5}_{-4.5}$  &  $\pm 1.7$ &        $38_{-4}^{+5}$ \\
523 & [ 486, 560]& $1935_{-408}^{+475}$ & $^{+6.5}_{-6}$  & $\pm 2.3$ &        $44_{-5}^{+5}$ \\
597 & [ 561, 635]& $1811_{-441}^{+511}$ & $^{+8}_{-7} $&  $\pm 3.0$  &        $43_{-6}^{+6}$ \\
671 & [ 636, 710]& $2100_{-546}^{+629}$ &  $^{+9.5}_{-8.5}$ &  $\pm 3.7$ &        $46_{-6}^{+6}$ \\
746 & [ 711, 785]& $2189_{-680}^{+777}$ &  $^{+11}_{-10}$ &  $\pm 4.6$ &        $47_{-8}^{+8}$ \\
856 & [ 786, 935]& $3104_{-738}^{+805}$ &  $^{+14}_{-12}$&  $\pm 5.6$ &        $56_{-7}^{+7}$ \\
1004 & [ 936, 1085]& $1084_{-1085}^{+1219}$ & $^{+18}_{-15}$ &  $\pm 7.7$  &    $33_{-22}^{+13}$ \\
1147 & [ 1086, 1235]& $223_{-2025}^{+2791}$ &  $^{+25}_{-18}$ &  $\pm 10.2$  &    $15_{-15}^{+29}$
\\
\hline
\end{tabular}
\end{center}
\caption{Uncorrelated Power Spectrum from the MAXIMA-1 map. Data for
$\ell < 335$ are from the power spectrum of the full map with 5\arcmin\
pixels, and data for $\ell > 335$ are from the power spectrum of the
central region of the map with 3\arcmin\ pixels.  Errors are the 68\%
integrated probability of the offset log-normal likelihood functions
with a constant prior in either $\ell(\ell+1)C_{\ell}/2\pi$ or $\Delta
T=\sqrt{\ell(\ell +1)C_{\ell}/2\pi} $ for their respective columns.
Here $[\ell_{min},\ell_{max}]$ gives the ranges of the dominant
bins. The correlated beam errors are described in the text.  }
\label{tab:cl_estimates}
\end{table*}

The top panel of Figure \ref{fig:power_spectrum} shows the maximum
likelihood power spectrum and an inflationary adiabatic model that
best fits the \citet{hanany_00} MAXIMA-1 and COBE/DMR power spectra.
The results for $\ell < 785$ are consistent with those in
\citet{hanany_00}.  The model has ($\Omega_{b}$, $\Omega_{cdm}$,
$\Omega_{\Lambda}$, $n$, $h$) = (0.07, 0.61, 0.23, 1, 0.60)
\citep{balbi_00}.  The $\chi^{2}$ for this model is 37 for all 41 data
points. If we only use the 13 data points of MAXIMA-1 we obtain
$\chi^2 = 5$ for this model.  In both cases, we are fitting with an
7 parameter inflationary model.

\section {Foregrounds}

Foreground sources include: emission from the earth, the atmosphere,
and galactic dust, free-free and synchrotron radiation, point sources,
and scattering due to the Sunyaev-Zel'dovich effect.  The analysis of
\citet{hanany_00} for $\ell < 785$ is applicable for sources that are
expected to be less strong at small angular scales.  In that work, we
argue that the detected signal is inconsistent with an atmospheric or
ground-based origin due to its temporal stability.  We can detect dust
contamination in our maps, but at a negligible level of 2.3 \microk\
at 150 GHz.  Estimates of bremsstrahlung and synchrotron radiation
\citep{bouchet_gispert_1999} predict contributions of less than 1
\microk\ at 150 and 240 GHz.  The ratio of detected power between 150
and 240 GHz is consistent with the CMB and inconsistent with dust,
synchrotron, and free-free emission.

Point sources are the foreground of concern for the high-$\ell$ region
data presented in this paper.  A catalog search
\citep{sokasian_etal2000,gawiser_smoot1997} yielded no detectable
radio or infra-red sources in any of the frequency bands.  The
``pessimistic'' model of \citet{tegmark_foreg} predicts $\sim 20$
\microk\ rms point source contamination at $\ell = 1000$, which is
small compared to our statistical errors.  The shape of the power
spectrum is inconsistent with a point source origin, which would give
a rising spectrum with increasing $\ell$.  Instead, the power spectrum
decreases above $\ell = 850$ and is consistent with zero.  A broad bin
covering $1235 < \ell < 1800$ (not tabulated) is consistent with zero
at the $1~\sigma$ level.

\section{Tests for Systematic Errors}

In the original data analysis of \citet{hanany_00}, we found no
evidence for uncorrected systematic errors over the range $35 < \ell <
785$ after performing a suite of tests.  Although maps with 8\arcmin\
and 10\arcmin\ resolution give power spectra with a systematic bias at
high $\ell$, they are still useful for systematic error tests especially in
the case where the result is a null power spectrum.  We confirm that
the following combinations of data produce a spectrum with no signal
over the range $35 < \ell < 1235$: (1) a dark bolometer, (2) the data
from the 410 GHz photometer, (3) the difference between the
overlapping part of the combined map from the CMB-1 and -2 scans, and
(4) the differences between the maps produced by different
photometers.  We repeated the fourth test on the central region of the
map using 3\arcmin\ pixels.  We calculated the difference of one and
the sum of the other two as shown in Fig.~\ref{fig:power_spectrum}

\section{Discussion}

We have presented a measurement of the angular power spectrum of the
CMB over a range of angular scales corresponding to the multipole
range $36 < \ell < 1235$, which is the largest yet reported by a
single experiment.  The data for $\ell > 400$ can be fit to a flat
line of amplitude 1860 \microk$^2$ with a $\chi^2 \sim 5$ for 7
d.o.f. However, if we focus on the bin at $\ell \sim 850$, we find
that its amplitude of $56 \pm 7$ \microk\ is in excess to the average
power of $42 \pm 3$ \microk\ in the range $441 < \ell < 785$ at a
confidence level of $\sim 95\%$.  This confidence level calculation is
presented in the companion paper by \citet{aardvark_00}.  This excess
power is consistent with the third acoustic peak predicted in this
region by inflationary adiabatic models.  The power spectrum is well
fit by such a model over the entire range of $\ell$.

\acknowledgments

We thank Danny Ball and the other staff at NASA's National Scientific
Balloon Facility in Palestine, TX for their outstanding support of the
MAXIMA program.  MAXIMA is supported by NASA Grants NAG5-3941,
NAG5-6552, NAG5-4454, GSRP-031, and GSRP-032, and by the NSF through
the Center for Particle Astrophysics at UC Berkeley, NSF cooperative
agreement AST-9120005, and KDI grant 9872979.  The data analysis used
resources of the National Energy Research Scientific Computing center
which is supported by the Office of Science of the U.S. Department of
Energy under contract no.\ DE-AC03-76SF00098, and the resources of the
Minnesota Supercomputing Institute.  PA acknowledges support from
PPARC rolling grant, UK.

\clearpage

\setcounter{figure}{0}

\begin{figure}
\plotone{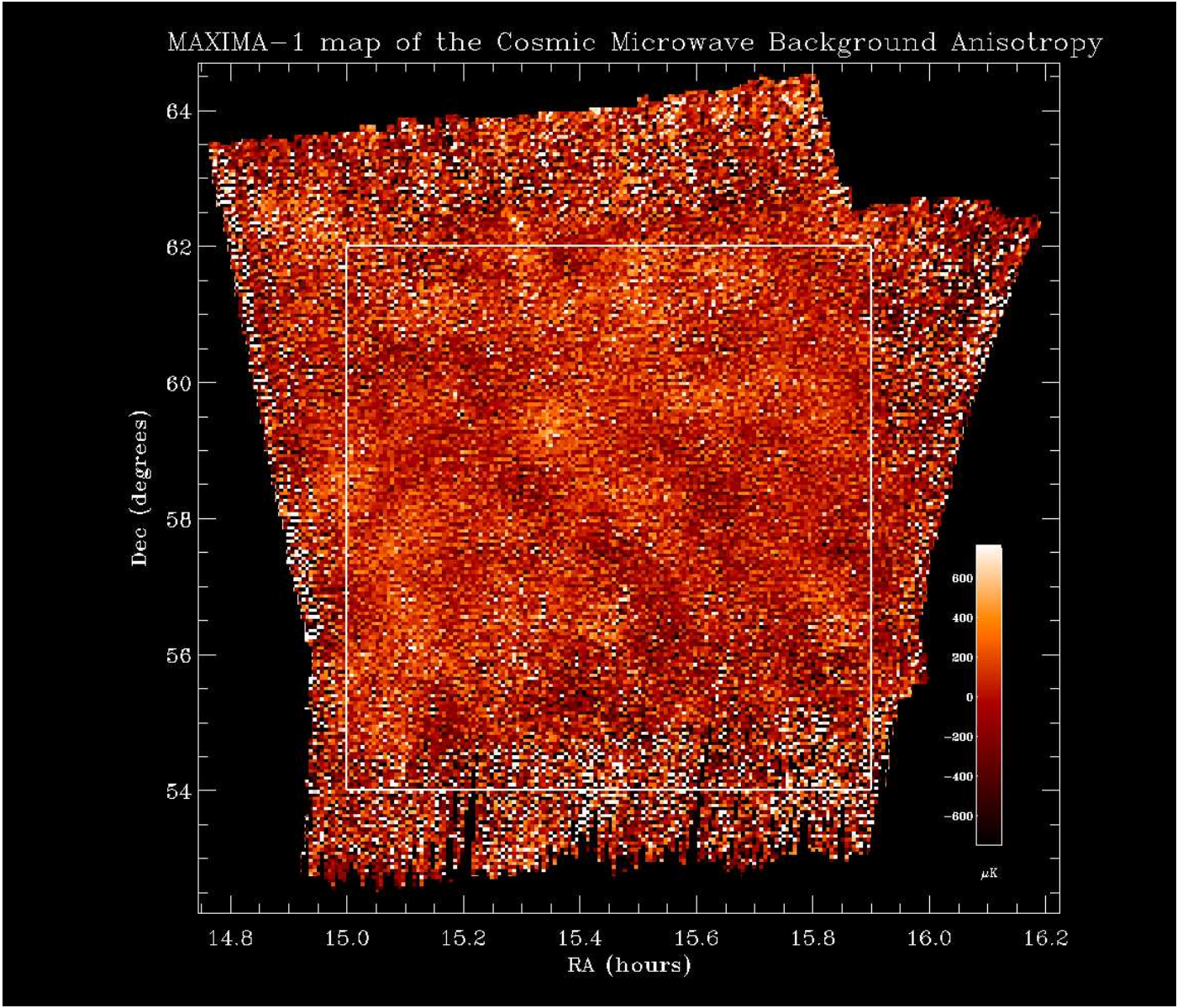}
\caption{A maximum-likelihood map of the CMB anisotropy from MAXIMA-1.
The resolution is determined by the 10\arcmin\ FWHM Gaussian beam of
the telescope.  The map is made using data from three 150 GHz
photometers and contains $\sim 40,000$ 3\arcmin$\times$3\arcmin\
pixels.  We use the central region of the map (outlined) to compute
the power spectrum presented in Fig. 2.  This region is fully cross
linked, has the most uniform sampling, and has the highest
signal-to-noise ratio.  The area of the central region is 60 deg$^2$
and contains $\sim 23,000$ pixels.}
\label{fig:maxima1_map}
\end{figure}

\begin{figure}
\centerline{\psfig{file=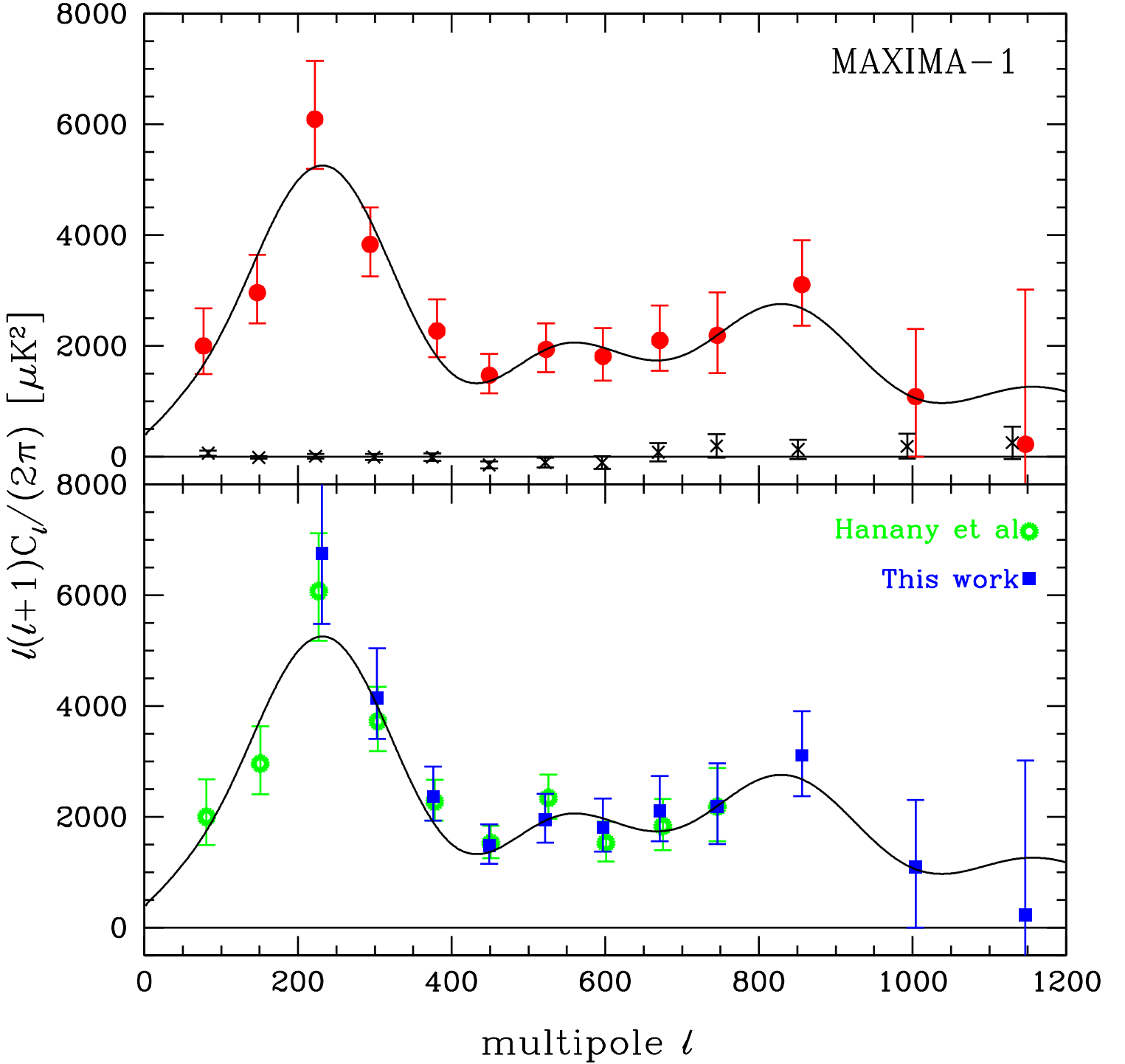} }
\caption{Top panel: Composite angular power spectrum of the CMB
anisotropy from the MAXIMA-1 map (filled circles).  The points for
$\ell < 335$ are from the power spectrum of the full 5\arcmin\
pixelized map from \citet{hanany_00}, and the points for $\ell > 335$
are from the power spectrum of the central region of the 3\arcmin\
pixelized map shown in Fig. 1. The error bars are 68\% confidence
intervals calculated using the offset log-normal likelihood functions
of \citet{bjk00}.  The solid curve is the best fit ($\Lambda$CDM)
inflationary adiabatic cosmology to the MAXIMA-1 and COBE/DMR
data. The model has ($\Omega_{b}$, $\Omega_{cdm}$, $\Omega_{\Lambda}$,
$n$, $h$) = (0.1, 0.6, 0.3, 1.08, 0.53) \citep{balbi_00}.  The crosses
are the estimated power spectrum of the difference between two
independent maps, the first given by one of the three photometers and
the other from the sum of the other two.  Bottom panel: A comparison
between the \citet{hanany_00} power spectrum and the power spectrum of
the central region of the 3\arcmin\ pixelized map shown in Fig. 1. For
the 3\arcmin\ central region data, the power spectrum for $\ell < 186$
is not well constrained.}
\label{fig:power_spectrum}
\end{figure}

\end{document}